\documentclass[aps,prb,twocolumn,showpacs]{revtex4}
\usepackage{graphics}
\usepackage{graphicx}
\usepackage{amssymb}
\begin{document}




\title{Multiferroic behavior in CdCr$_2X_4$ ($X=$~S, Se)}
%

\author{J. Hemberger$^{1,2}$, P. Lunkenheimer$^{2}$, R. Fichtl$^{2}$, S. Weber$^{2}$,
V. Tsurkan$^{2,3}$, A. Loidl$^{2}$}

\address{
 $^{1}$II. Physikalisches Institut, Universit\"at zu
 K\"oln, 
 D-50937 K\"oln, Germany \\
 $^{2}$
 Center for Electronic Correlations and Magnetism, University of Augsburg, D-86135
Augsburg, Germany \\
 $^{3}$Institute of Applied Physics, Academy of Sciences of Moldova,
 MD 2028, Chisinau, Republic of Moldova }


\begin{abstract}

The recently discovered multiferroic material CdCr$_2$S$_4$ shows a
coexistence of ferromagnetism and relaxor ferroelectricity together
with a colossal magnetocapacitive effect. The complex dielectric
permittivity of this compound and of the structurally related
CdCr$_2$Se$_4$ was studied by means of broadband dielectric
spectroscopy using different electrode materials. The observed
magnetocapacitive coupling at the magnetic transition is driven by
enormous changes of the relaxation dynamics induced by the
development of magnetic order.

\end{abstract}




%

%
%

\maketitle

\begin{figure}[!ht]
\begin{center}
\includegraphics[width=0.5\textwidth]{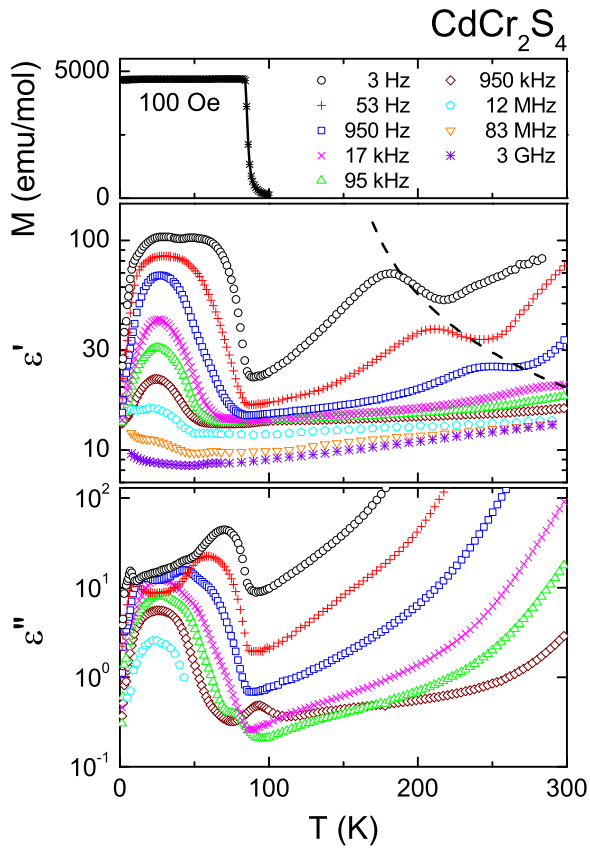}
\end{center}
\caption{Temperature dependence of the low field magnetization
(upper frame) and complex dielectric permittivity $\varepsilon^*$
(lower frames) of CdCr$_2$S$_4$ for various frequencies. The
magnetic transition occurs at $T\approx 84$~K and is accompanied
by an increase of the permittivity towards low temperatures.
(Note, that the plateau-like low temperature value of the
magnetization is determined by the demagnetization factor of the
sample in the small measurement field of $H=100$~Oe and does not
display an intrinsic saturation.)} \label{ccs}
\end{figure}
\begin{figure}[!ht]
\begin{center}
\includegraphics[width=0.5\textwidth]{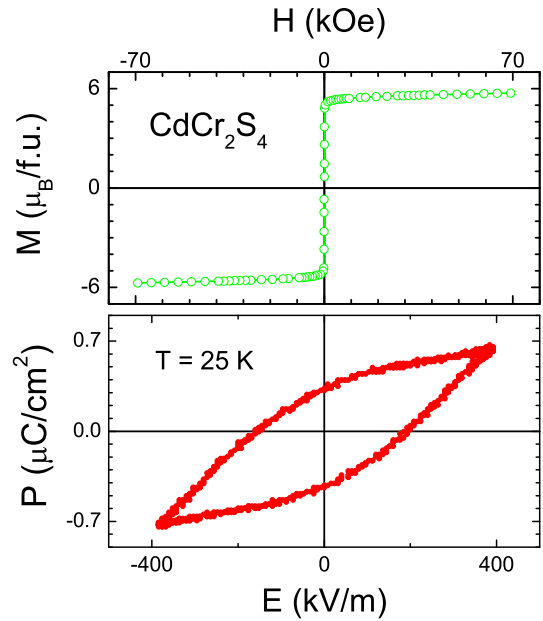}
\end{center}
\caption{Hysteresis loops for the magnetic ($M(H)$, upper frame)
and dielectric sector ($P(E)$, lower frame) measured at $T=25$~K.
The magnetic data was corrected for demagnetization effects; the
polarization data was measured at $\nu=1.13$~Hz.}
\label{hysteresis}
\end{figure}

\begin{figure}[!ht]
\begin{center}
\includegraphics[width=0.5\textwidth]{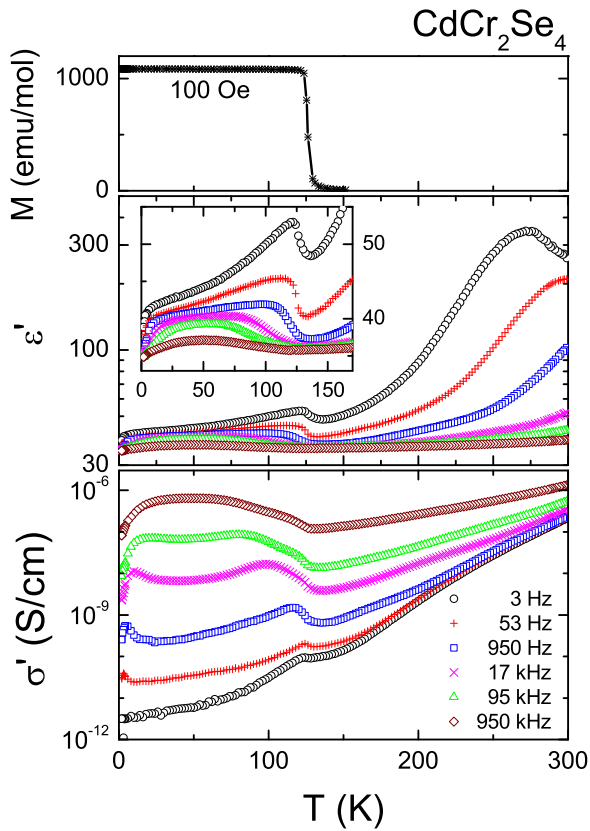}
\end{center}
\caption{Temperature dependence of the low field magnetization
(upper frame), real part of the dielectric permittivity
$\varepsilon'$ (middle), and real part of the conductivity (lower
frame) of CdCr$_2$Se$_4$ for various frequencies. The magnetic
transition occurs at $T\approx 125$~K and is accompanied by an
increase of the permittivity towards low temperatures. (Note, that
the plateau-like low temperature value of the magnetization is
determined by the demagnetization factor of the sample in the small
measurement field of $H=100$~Oe and does not display an intrinsic
saturation.)} \label{ccse}
\end{figure}
\begin{figure}[!ht]
\begin{center}
\includegraphics[width=0.5\textwidth]{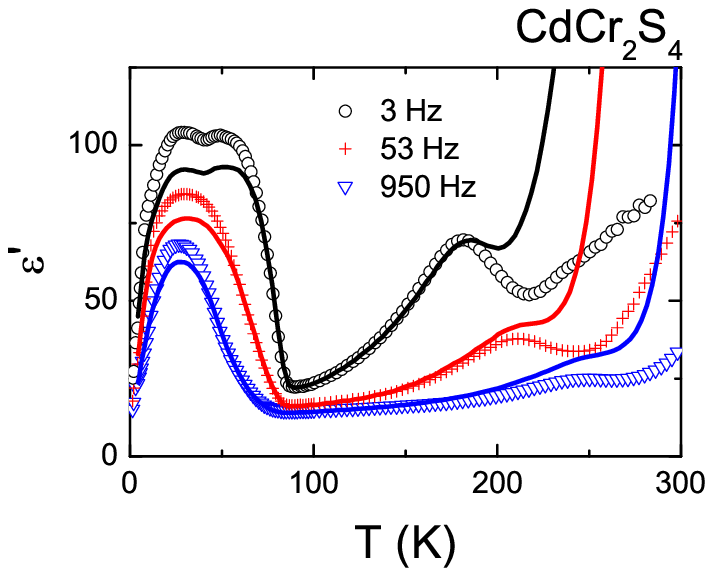}
\end{center}
\caption{Temperature dependence of the real part of the dielectric
permittivity $\varepsilon'$ of CdCr$_2$S$_4$ measured for various
frequencies. The symbols represent measurements employing
sputtered gold electrodes; the lines denote the results obtained
with electrodes of silver paint.} \label{contacts}
\end{figure}

In multiferroic materials at least two (in a strict sense)
ferro-type order parameters corresponding to different microscopic
degrees of freedom coexist simultaneously. Such a scenario can
combine, e.g., ferro-orbital order, ferroelasticity,
ferromagnetism, or ferroelectricity. Especially the coexistence of
the latter properties, namely ferroelectricity and ferromagnetism,
is rarely found and represents a "hot" topic in recent solid state
research \cite{1,2,hemberger}. The possible coupling of both
sectors, i.e.\ the strong variation of electric (magnetic)
properties under application of a magnetic (electric) field, which
is found in some of these materials, makes them highly attractive
not only from an academic point of view, but also for potential
applications in microelectronics. In the present contribution we
report on magnetization, dielectric polarization, as well as
broadband dielectric measurements in the multiferroic systems
CdCr$_2$S$_4$ and CdCr$_2$Se$_4$ and provide detailed information
on the dielectric relaxation dynamics in the paramagnetic and
ferromagnetic state.
For these investigations special emphasis was put on the influence
of contacts on the dielectric characterization comparing sputtered
gold- and silver-paint electrodes applied on opposite sides of the
plate-like single crystalline samples. Dielectric constant and
loss were measured over a broad frequency range (3 Hz $<  \nu  <
3$~GHz) using frequency-response analysis and a reflectometric
technique \cite{lunki-mess}. Electric polarization was detected
employing a high-impedance Sawyer-Tower circuit and the
magnetization data was recorded with a commercial SQUID
magnetometer (MPMS, QuantumDesign).


At room temperature, the recently discovered multiferroic
CdCr$_2$S$_4$ is a cubic spinel compound \cite{hemberger}. The
Cd$^{2+}$ ions on the structural $A$-sites carry neither a
magnetic nor an orbital degree of freedom. The Cr$^{3+}$ ions on
the octahedrally surrounded $B$-sites possess half filled $t_{2g}$
shells and thus also are orbitally inactive but due to Hund's
coupling and a quenched orbital moment carry a $S=J=3/2$ spin
configuration. It shall be noted, that the absence of a
Jahn-Teller active orbital degree of freedom and the high
crystallographic symmetry allow for a relaxation of the structure
into polar distortions. The spins are coupled ferromagnetically
and as can be seen in the upper frame of Fig.~\ref{ccs} magnetic
order sets in at $T_c\approx 84$~K. At the same time a steep
increase of the dielectric constant can be detected, which for low
frequencies reaches relatively high values above
$\varepsilon'\approx 100$ (Fig. 1, middle frame). The details of
the frequency dependence of the complex permittivity will be
discussed below, but before it shall be pointed out that the low
temperature phase of CdCr$_2$S$_4$ exhibits not only spontaneous
magnetization but in addition the high values of $\varepsilon'$
can be interpreted as precursor of the onset of remnant dielectric
polarization at lower temperatures. Fig.~\ref{hysteresis} shows
hysteresis loops taken at $T=25$~K for both, the magnetic as well
as the dielectric sector. $M(H)$ shows the characteristics of a
soft ferromagnet reaching nearly the full magnetic moment of
$6\mu_B$ expected for two Cr$^{3+}$ ions per formula unit. The
small deviations from the full value can be attributed to
pronounced magnon features typical for such systems. At the same
time the $P(E)$ curve is quite slim and smeared out. No real
saturation value is reached within the experimentally accessible
electric field range of 400~kV/m and no well defined coercive
electric field strength can be identified. Such type of hysteresis
behavior is typical for so-called relaxor-ferroelectric behavior,
where nano-scale ferroelectric clusters determine the polarization
response \cite{cross}.

Such type of behavior is also evidenced by the frequency and
temperature dependence of the complex permittivity in
CdCr$_2$S$_4$. Fig.~\ref{ccs} shows real and imaginary part of the
dielectric constant, $\varepsilon'(T)$ and $\varepsilon''(T)$, for
frequencies from 3 Hz to 3 GHz. For $T > 100$~K, in
$\varepsilon'(T)$ a peak shifting to lower temperatures and
increasing in amplitude with decreasing frequency shows up. The
dashed line indicates a Curie-Weiss law,
$\propto 1/(T-135$~K$)$, for the right flank of the peaks, which
can be taken as an estimate of the static dielectric constant. The
further increase of $\varepsilon'(T)$ towards higher temperatures
is due to contact and conductivity contributions as discussed
below. Thus the overall characteristics of the relaxational
behavior in CdCr$_2$S$_4$ resembles that observed in relaxor
ferroelectrics \cite{cross}. There the reduction of $\varepsilon'$
below the peak temperature is usually ascribed to a cooperative
freezing-in of ferroelectric clusters on the time scale given by
the frequency of the applied AC electric-field, quite in contrast
to canonical ferroelectrics where the frequency dependence of
$\varepsilon'(T)$ is negligible. Such relaxational features seen
in the real part of the permittivity should be accompanied by
peaks in the imagninary part. At $T > T_c$ in CdCr$_2$S$_4$, they
should appear at the frequency of the point of inflection at the
left wing of the maxima in $\varepsilon'(T)$. However, no such
peaks become obvious in Fig.~\ref{ccs}, which can be ascribed to
superimposing contributions from charge carrier transport. As the
dielectric loss and the real part of the conductivity are linked
via the relation $\varepsilon''\propto \sigma'/\nu$ the staggered
behavior of the dielectric loss curves can be attributed to the
presence of conductivity contributions.

The most remarkable feature in Fig.~\ref{ccs} is the strong
increase of $\varepsilon'(T)$ below the ferromagnetic transition
temperature $T_c = 84$~K, indicating the close coupling of
magnetic and dielectric properties. It can be explained assuming
that the frozen-in dynamics of the dielectric relaxor entities are
melting due to the onset of magnetization and become fast again,
which restores a large contribution in the dielectric response
\cite{ccsPRL}. At the same time the dielectric loss is increasing
below $T_c$. As mentioned before, this in part may be influenced
by changes in the conductivity. Magnetoresistive effects have been
reported for CdCr$_2$S$_4$ and CdCr$_2$Se$_4$ in literature
\cite{lehmann}.

For CdCr$_2$Se$_4$ the magnetic transition is shifted to higher
temperatures. The magnetic order sets in at $T_c \approx 125$~K as
monitored in the upper frame of Fig.~\ref{ccse}. At this
temperature an increase of $\varepsilon'(T)$ towards lower
temperatures is detected, too. Thus, similar to CdCr$_2$S$_4$
\cite{hemberger}, the closely related CdCr$_2$Se$_4$ also exhibits
a strong magneto-dielectric coupling. The typical relaxor peaks in
the real part of the permittivity at $T > T_{c}$ are shifted to
higher temperatures and only can be detected for the lowest
frequencies within the examined temperature regime below 300~K.
The superimposed influence of conductivity is higher in
CdCr$_2$Se$_4$ compared to CdCr$_2$S$_4$. For this in the lowest
frame of Fig.~\ref{ccse} $\sigma'$ is plotted. For low frequencies
the data falls onto one curve reflecting the DC value of
conductivity. The higher-frequency curves successively branch off
from this curve pointing to additional relaxational and/or ac
conductivity contributions.

It has to be mentioned that even though the observed effects are
stable against the exchange of the chalcogen ion, the occurence of
the observed dielectric features depend sensitively on the
stoichiometry. Annealing of the samples, both, in vacuum or
sulphur/selen atmosphere, leads to a suppression of the frequency
depend relaxor peaks in $\varepsilon'(T)$ above $T_c$ and no
remanent electric polarization can be found at low temperatures.
Also, these features so far could not be detected in polycrystalline
samples.



Also, one should consider the possibility that the observed
relaxation features are not a bulk property, but of the so-called
Maxwell-Wagner type \cite{maxwell-wagner}, i.e. caused by
polarization effects at or close to the surface of the samples. A
prominent example for Maxwell-Wagner relaxations are the so-called
"colossal dielectric constant" materials \cite{peters-antihighperm}.
In most (if not all) of these materials, relaxational behavior with
very high (typically $10^3$-$10^5$), non-intrinsic dielectric
constants arise from the formation of depletion layers at the
interface between sample and metallic contacts. The high capacitance
of these insulating layers at the sample surface can lead to
apparently high dielectric constants and relaxational behavior. Thus
in the present case, we varied the electrode materials, to check for
such a scenario. Fig.~\ref{contacts} shows measurements of
$\varepsilon'(T)$ for contacts made from silver-paint or sputtered
gold. For high temperatures significant differences between both
measurements can be detected. However, the relaxation features are
still present at the same temperatures and for temperatures below
the data match each other. This clearly points towards a dominance
of intrinsic effects. 

Concerning the microscopic origin of the polar moments in
CdCr$_2$S$_4$ and CdCr$_2$Se$_4$ it can be assumed that the
ferroelectric distortions result from an off-center position of
the Cr$^{3+}$-ions which generates a locally polar but
macroscopically isotropic cluster state \cite{grimes}. Geometrical
frustration within the at room temperature highly symmetric cubic
lattice drives the observed relaxor-like freezing \cite{ramirez}.
On the other hand, the origin of the strong coupling of
magnetization and dielectric permittivity in this compound is so
far unknown.
As pointed out the relaxation dynamics of the polar moments are
accelerated below $T_c$ \cite{ccsPRL}, but it remains to be
clarified what is the microscopic origin of the detected
relaxation dynamics and why this dynamics couples so strongly to
the magnetic order parameter. A coupling via exchangestriction,
i.e. volume changes arising from the magnetic exchange energy
\cite{martin} seems possible. The onset of spin order leads to a
softening of the lattice thereby reducing the energy barriers
against dipolar reorientation and thus enhancing the mean
relaxation rate. As an alternative explanation one could consider
a magnetic-field induced variation of charge-carrier mobility or
density. As mentioned before, a sizable DC magnetoresistive effect
is well known for CdCr$_2$S$_4$ \cite{lehmann}, but it cannot be
responsible for the observed anomalies of  $\varepsilon'$ as the
DC resistivity only contributes to $\varepsilon''$ In contrast,
hopping-type charge transport is known to give rise to frequency
dependent AC conductivity, which via the Kramers-Kronig relation
would lead to a contribution to the dielectric constant
\cite{jonscher}.


In conclusion, our results clearly reveal that in CdCr$_2$S$_4$
and CdCr$_2$Se$_4$ canonical ferromagnetism coexists at sizable
ordering temperatures with a relaxor-ferroelectric state,
characterized by a significant relaxational behavior. Both order
parameters are strongly coupled. There is a radical change of the
dielectric relaxational dynamics driven by the onset of
magnetization that leads to the observed strong increase of the
dielectric permittivity in these compounds. The present dielectric
experiments cannot provide final evidence on the microscopic
origin of this puzzling behavior. While contact effects could be
ruled out and the influence of charge transport seems unlikely, a
scenario in which the relaxation mechanism interacts with magnetic
order via exchangestriction can be considered the most plausible.

\section*{Acknowledgement}
This work was partly supported by the Deutsche
Forschungsgemeinschaft via the Sonderforschungsbereich 484 and
partly by the BMBF via VDI/EKM, FKZ 13N6917.

\bibliographystyle{prsty}

\end{document}